\documentclass[a4paper,11pt]{article}

\usepackage{amssymb,amstext,amsmath,amsthm}
\usepackage{graphicx}
\usepackage{latexsym}
\usepackage{psfrag}
\usepackage{amsfonts}
\usepackage{vmargin}

\newcommand{\bea}{\begin{eqnarray}}
\newcommand{\eea}{\end{eqnarray}}
\newcommand{\nn}{\nonumber}
\newcommand{\be}{\begin{equation}}
\newcommand{\ee}{\end{equation}}

\newtheorem{theo}{Theorem}
\newtheorem{lem}{Lemma}

\newcommand{\R}{\mathbb{R}}
\newcommand{\C}{\mathbb{C}}

\newcommand{\ftj}{\mbox{15}j}   
\newcommand{\lalg}[1]{\mathfrak{#1}}  
\newcommand{\SU}{\mathrm{SU}}
\newcommand{\SO}{\mathrm{SO}}
\newcommand{\SL}{\mathrm{SL}}

\newcommand{\su}{\lalg{su}}

\newcommand{\dd}{\mathrm{d}}

\newcommand{\nb}{\mathbf{n}} 
\newcommand{\bb}{\mathbf{b}} 

\def\ra{\rangle}

\def\bra#1{\mathinner{\langle{#1}|}}
\def\ket#1{\mathinner{|{#1}\rangle}}

\setmarginsrb{2.5cm}{1cm}{2.5cm}{2.5cm}{12pt}{11mm}{0pt}{13mm}

\title{Quantum gravity asymptotics from the $\SU(2)$ $\ftj$ symbol}
\author{John W. Barrett$^{a}$\footnote{john.barrett@nottingham.ac.uk} \, , \, Winston J. Fairbairn$^{a,b}$\footnote{winston.fairbairn@uni-hamburg.de} \, , \, Frank Hellmann$^{a}$\footnote{frank.hellmann@maths.nottingham.ac.uk}
\vspace{2mm}
\\ [1mm]
\itshape{\normalsize{$^a$School of Mathematical Sciences, University of Nottingham,}} \\
\itshape{\normalsize{University Park, Nottingham NG7 2RD,}} \\
\itshape{\normalsize{UK}}
\\
\\
\itshape{\normalsize{$^b$Department Mathematik, Universit\"at Hamburg,}} \\
\itshape{\normalsize{Bundesstrasse 55,  20146 Hamburg,}} \\
\itshape{\normalsize{Germany}}}
\date{December 24th, 2009}
\begin{document}

\maketitle

\vspace{-3mm}
\begin{abstract}
The asymptotics of the SU(2) 15j symbol are obtained using coherent states for the boundary data. The geometry of all non-suppressed boundary data is given. For some boundary data, the resulting formula is interpreted in terms of the Regge action of the geometry of a 4-simplex in 4-dimensional Euclidean space.
This asymptotic formula can be used to derive and extend the asymptotics of the spin foam amplitudes for quantum gravity models. The relation of the SU(2) Ooguri model to these quantum gravity models and their continuum Lagrangians is discussed.
  \end{abstract}

\section{Introduction}

Spin foam models for quantum gravity are state sum models based on spin networks of various kinds. In these models, the variables are representations and intertwiners of a group (or quantum group) labelling the edges and vertices of the spin networks. The amplitude for a simplex is defined by gluing these according to the combinatorics of the simplex, to make a closed spin network. The first model of this type was the Ponzano-Regge model, for three-dimensional space-times \cite{PR}, where the 3-simplex amplitude is a $6j$-symbol. Models in this spirit were developed for four-dimensional space-times in \cite{Barrett2000,Barrett1998,Engle2008b,Engle2008,Engle2007,Pereira2008,Livine2008,Freidel2008}. These can be viewed as modifications or constrained versions of the much simpler Ooguri model \cite{Ooguri1992c,Crane1993} corresponding to a quantisation of BF theory. This topological model is based on the $\SU(2)$ $\ftj$-symbol.

A first insight into the geometric interpretation underlying these models can be seen in the large spin behaviour of the amplitude for a single 4-simplex. The intertwiner degrees of freedom can be identified with the shape space of tetrahedra with areas given by the spins \cite{Barbieri:1997ks,Baez:1999tk,Freidel2009,Conrady2009}. 
It was shown in \cite{Barrett2009} that if the data is such that these tetrahedra, as a geometric boundary, induce a non-degenerate Euclidean metric on the 4-simplex, the large spin asymptotics of the 4-simplex amplitude of the gravity models with Immirzi parameter\cite{Engle2008b,Engle2008,Engle2007,Pereira2008,Livine2008,Freidel2008} can be understood in terms of this geometry. In particular the phase part of the asymptotic behaviour is the Einstein action associated to this geometry in Regge calculus \cite{Regge1961}, as is expected in a discrete quantization of general relativity and as is the case, for example, in the Ponzano Regge model \cite{PR}. In this analysis, four terms were found to contribute to the asymptotics with phases proportional to the Einstein action. While it was expected to find two contributions from the two orientations associated to this boundary data, the existence of additional terms came as a surprise. 

In this paper we apply and extend the methods of \cite{Barrett2009} to study the asymptotics of the $\SU(2)$ $\ftj$-symbol underlying the new models. We find that the asymptotics are governed by the expected classical solutions of the topological theory. We find however that for some boundary data these solutions retain an interpretation in terms of $4$-dimensional Euclidean geometry and the phase part of the asymptotics of the $\ftj$-symbol is given by the Regge action. This provides a proof for a part of the asymptotic formula for the the Lorentzian model \cite{Barrett2009a} that was missing. The obtained asymptotics for the $\ftj$-symbol are then used to derive the asymptotic formulae for the four-simplex amplitudes of the new Euclidean gravity models expressed as rescaled squares of the $\ftj$-symbol. This provides a nice explanation for the appearance of the weird terms in \cite{Barrett2009}. A further explanation for the non-geometrical terms in the asymptotics of the gravity models is explained by the fact that the gravity models appear to contain a single copy of the Ooguri model as a subset of the configurations. This interference between the topological Ooguri model and the gravity configurations has some implications for the continuum limit of the gravity state sum models, as discussed in section \ref{discussion}.
Finally, there is a heuristic interpretation of our results in terms of the corresponding Lagrangian continuum field theories.


\section{$\SU(2)$ $\ftj$-symbol asymptotics}

The $\ftj$-symbol is determined by data associated to the boundary of a 4-simplex. This data is a spin $k$ in $\mathbb{N}/2$ for each triangle and a four-valent intertwining operator for each tetrahedron. By introducing coherent states for the spins $k$,  one can represent each intertwiner in the coherent state basis \cite{Livine2007a} by $\SU(2)$-averaging the tensor product of four coherent states. The notation used is as follows. A coherent state for the spin $\frac12$ representation is labelled by a unit vector $\mathbf n$ in $\R^3$, and is denoted $\ket \nb$; the phase of this coherent state is arbitrary at present and will be specified later. 
The tetrahedra in the 4-simplex are labelled with $a=1,2,\ldots,5$. The spins label the triangles of the 4-simplex and so are specified by $k_{ab}=k_{ba}$, for each $a\ne b$. 

The intertwiner for the $a$-th tetrahedron is given by the formula
$$\psi_a=\int_{X\in\SU(2)}\dd X\;\bigotimes_{b\ne a} X \ket{\nb_{ab}}^{2 k_{ab}},$$
and the boundary state for the whole simplex is 
$$\psi=\bigotimes_a \psi_a.$$
The data $\{k_{ab},\nb_{ab}\}$ specifying this boundary state up to phase is called the boundary data. The phase choice will be discussed later.

The $\ftj$ symbol in the coherent state basis is given by the following integral formula \cite{Livine2007a}, \cite{Barrett2009}
\be
\ftj(k,\nb) = (-1)^\chi \int_{\SU(2)^5} \prod_{a=1..5} \dd X_a \prod_{1\leq a < b \leq 5} \bra{J \nb_{ab}} X^\dagger_a X_b \ket{\nb_{ba}}^{2k_{ab}}.
\ee
Here,   $\mathstrut\dagger$ is the Hermitian conjugate,  $J$ is the standard anti-linear $\SU(2)$ operator and the notation $\bra{J \nb_{ab}}$ stands for the covector dual to $J\ket{\nb_{ab}}$. Finally, the sign factor $(-1)^\chi$ is determined by the graphical calculus relating the $\ftj$ spin network diagram to the above evaluation. The formula is a linear function of the boundary state $\psi$ and is determined by contractions, according to the combinatorics of the diagram, using the bilinear inner product on spin $k$ representations.

\subsection{Asymptotic problem and critical point equations}

The above formula is an integral expression in exponential form and so the asymptotic limit, where all spins are simultaneously rescaled, $k_{ab} \rightarrow \lambda k_{ab}$, and taken to be large ($\lambda \rightarrow \infty$), can be analysed with stationary phase methods. 

The action for the asymptotic problem yields
\be
\label{action}
S_{(k,\nb)}[X] = \sum_{a < b} 2 k_{ab} \ln \, \bra{J \nb_{ab}} X^\dagger_a X_b \ket{\nb_{ba}}.
\ee
It enjoys a global $\SU(2)$ continuous symmetry and a discrete $\pm$ symmetry at each vertex $a$, given by the formula
\begin{equation}\label{symmetry} X_a'=\epsilon_aYX_a,\end{equation}
with $Y\in\SU(2)$ and $\epsilon_a=\pm1$.

The critical points of this action satisfy the same equations as in \cite{Barrett2009}. The stationarity with respect to the group variables implies closure for each tetrahedron $a$
\be
\label{closure}
\sum_{b: b\ne a} k_{ab}\nb_{ab}=0,
\ee
and the critical points are those which also satisfy
\be
\label{equations}
 X_b \nb_{ba} = - X_a\nb_{ab}
\ee
for each triangle $ab$, $a \neq b$.

\subsection{Solutions to the critical point equations}

In this section, we firstly show that the solutions to the above critical point equations
admit an interpretation in terms of $BF$ theory. Then, we study the classification problem of these solutions for various classes of boundary data.

\subsubsection{Geometry of the solutions: relation to $BF$ theory}

The solutions to the critical point equations \eqref{closure} and \eqref{equations} can be interpreted as the solutions of a four-dimensional $BF$ theory with group $\SU(2)$ discretised on a $4$-simplex. The fundamental fields of the theory are an $\SU(2)$ connection $A$ and an $\su(2)$-valued two-form $B$. The equations of motion of the theory are solved by covariantly closed $B$ fields and flat connections:
$$
\dd_A B=0, \;\;\;\;\; \mbox{and} \;\;\;\;\; F_A=0,
$$
where $F_A$ is the curvature of $A$.
Note that since flat connections are locally pure gauge, the first equations reduces locally to $\dd B=0$.

Evaluating the $BF$ action requires an orientation of the 4-manifold. In the discrete setting, an orientation of a 4-simplex establishes a coherent rule for associating an orientation to the triangles. Triangle $t_{ab}$ is defined to be the intersection of tetrahedra $a$ and $b$ with the orientation induced from the boundary of tetrahedron $a$, which itself inherits an orientation from the 4-simplex. Thus, we have that $t_{ba}=-t_{ab}$.

The analogy between $BF$ theory and the asymptotic data is apparent by defining the following variables
$$\bb_{ab}=k_{ab}X_a\nb_{ab}.$$
The solutions to the critical point equations can then be parameterised by the $X_a$ and $\bb_{ab}$. These are the discrete connection and $B$-field variables, respectively. This interpretation follows from the critical point equations \eqref{closure} and \eqref{equations}, which expressed in terms of the $\bb_{ab}$, read
\begin{equation}
\label{vequations}
\sum_{b:b\ne a}\bb_{ab}=0,\qquad\qquad \bb_{ab}=-\bb_{ba}.
\end{equation}
A such set of twenty three-dimensional vectors $\bb_{ab}$ satisfying the above two vector equations determines a geometrical structure called a {\em vector geometry}. Accordingly, a vector geometry is parametrised by $20 \times 3 - (4 \times 3 + 10 \times 3) = 18$ numbers.
This data determines a constant $B$-field on the 4-simplex, namely the 2-form valued in the Lie algebra $\su(2)$ with coefficients which are constant in the standard linear coordinates on the simplex, satisfying
$$\int_{t_{ab}} B=\bb_{ab}.$$
The closure equation $\sum \bb_{ab}=0$ is now interpreted as Stokes' theorem for the $B$ field around the boundary of a tetrahedron, that is, the first equation of motion of $BF$ theory $dB = 0$.

The proof that the existence of the constant $B$-field is equivalent to \eqref{vequations} is as follows. Pick a distinguished vertex of the 4-simplex and denote the four edge vectors pointing away from that vertex by $\xi_1,\ldots,\xi_4$. A basis for the space of bivectors $\Lambda^2(\R^4)$ is given by $\xi_p\wedge\xi_q$. Therefore the components of $B$ with respect to this basis are $B( \xi_p\wedge\xi_q)$ which are just the integrals of $B$ on the triangles meeting the distinguished vertex. Using the closure relation shows that these are all the independent variables in a solution to \eqref{vequations}.
 
The $X_a$ variables are a discrete version of the connection. This is because the gluing of two 4-simplexes proceeds by the identification of the $\nb_{ab}$ variables on a common tetrahedron. In terms of the $\bb_{ab}$ variables, this means the gluing takes place \emph{after} the action of the corresponding $X_a$ for the tetrahedron. The $X_a$ is therefore a parallel transport operator for half the dual edge between the two 4-simplexes. As the boundary data has not been varied in the critical equations, the flatness equation of motion for this connection is not yet enforced because it is invisible at the level of a single $4$-simplex. This equation appears only when 4-simplexes are glued all the way around a triangle. 

\subsubsection{Classification of the solutions}

The existence and a partial classification of the solutions to the critical point equations \eqref{closure} and \eqref{equations} was established in \cite{Barrett2009} and depends on the boundary data. 
Two solutions $X_a$, $X_a'$ are said to be equivalent if they are related by the symmetry operation \eqref{symmetry} and inequivalent otherwise. In theorem \ref{classification} the results of \cite{Barrett2009} are extended  by studying the existence and classification of these equivalence classes of solutions as a function of the boundary data.

\paragraph{General boundaries.}

For general boundary data, the solutions to the vector geometry equations \eqref{vequations}, for fixed $|\bb_{ab}|$, were analyzed by Barrett and Steele \cite{Barrett:2002ur}. The vectors for a given tetrahedron describe a quadrilateral in $\R^3$. Thus there are no such solutions unless the  $|\bb_{ab}|$ satisfy the inequalities of a quadrilateral (i.e., each one can be no larger than the sum of the other three). Given a solution to \eqref{vequations}, one can glue up an open cubical box starting at any quadrilateral. The equations imply that the 5 quadrilaterals glue up as 5 sides of a box and the open side of the box is a parallelogram. Conversely any such box geometry determines solutions to \eqref{equations}. Barrett and Steele showed that for a given set of $k_{ab}$, determining the edge lengths of the box, there is a 5-dimensional set of such box geometries. 

\paragraph{Non-degenerate boundaries.}
There is an important subset of boundaries where the above results also apply.
Boundary data such that for each tetrahedron $a$, the face vectors $\nb_{ab}$, for fixed $a$ and varying $b$, span a three-dimensional space is called a {\em non-degenerate} boundary data. In this case, if the four vectors $\nb_{ab}$ satisfy the closure condition \eqref{closure} they specify an embedding of the tetrahedron in three-dimensional Euclidean space, such that the vectors are the outward face normals and the $k_{ab}$ are the areas. In this way, each tetrahedron inherits a metric and an orientation but the metrics and orientations of different tetrahedra do not necessarily match. 
For non-matching boundary data, the classification established in \cite{Barrett2009} states that there is, up to equivalence, one or no solution to the critical point equations.

\paragraph{Geometric boundaries.}\label{reggelike}

Non-degenerate boundary data for the whole $4$-simplex is said to be {\em geometric} or {\em Regge-like} if the closure constraint \eqref{closure} is satisfied and if the individual tetrahedron metrics and orientations glue together consistently to form an oriented Regge-calculus 3-geometry for the boundary of the 4-simplex. This is the requirement that the induced metrics on the triangles agree for both of the tetrahedra sharing any given triangle, and the induced orientations are opposite. This boundary data satisfies the gluing constraints discussed by Dittrich, Ryan and Speziale in \cite{Dittrich2008, Dittrich2008a}, which were found there to be necessary to ensure geometricity.

For geometric boundary data, there exists a unique (up to a $\mathbb{Z}_2$ lift ambiguity discussed in \cite{Barrett2009}) set of ten $\SU(2)$ elements $g_{ab}= g_{ba}^{-1}$ which glue together the oriented geometric tetrahedra of the boundary, defined as follows.  The embedding of tetrahedron $a$ in Euclidean space determines a position vector $\mathbf x_a^v$ for vertex $v$. Suppose $v$ and $w$ are vertices which appear in both tetrahedron $a$ and tetrahedron $b$. Then $g_{ba}$ is a spin lift of the unique rotation which maps $\mathbf x_a^v- \mathbf x_a^w$ to $\mathbf x_b^v- \mathbf x_b^w$, for each possible choice of $v$ and $w$, and maps the outward normal to one tetrahedron to the inward normal to the other, 
\be \label{gluevector} g_{ba} \, \nb_{ab} = - \nb_{ba}.\ee
This gluing map can also be interpreted as the parallel transport operator $g_{ba} : T_a \rightarrow T_b$ mapping the tangent space $T_a$ of the $a$-th tetrahedron to the tangent space $T_b$ of the $b$-th tetrahedron%
\footnote{Note that to account for this our convention here differs by $g_{ab} \rightarrow g_{ba}$ from \cite{Barrett2009}. The vector $\nb_{ab}$ is interpreted to be in tetrahedron $a$. With this convention $g_{ba}$ goes from $a$ to $b$ which is notationally more convenient and in accordance with the parallel transport conventions.}
 and describing the change of reference frame from $a$ to $b$.

From this data, one can make a canonical choice of phase for the boundary state $\psi$ by picking the phases of the coherent states such that
\be \label{ReggeState}
| \mathbf n_{ba} \ra =  g_{ba} J|  \mathbf n_{ab} \ra.
\ee
The boundary state $\psi$ with this choice of phase is called a {\em Regge state}.


Geometric boundaries of a $4$-simplex fall in three geometrical classes, depending on whether it is the data of a $4$-simplex with 4d Lorentzian (with space-like tetrahedra), 4d Euclidean or 3d Euclidean geometry. This follows from the fact that a metric for the 4-simplex is determined uniquely by the edge lengths. Since the three-dimensional hypersurfaces containing the tetrahedra all have signature $+++$, the signature for this 4-metric can only be $-+++$, $++++$ or $0+++$. In the last case, the metric is a pull-back of the Euclidean metric on $\R^3$ by a projection. To complete the description of geometric boundary data initiated in \cite{Barrett2009} and \cite{Barrett2009a}, where the boundary data for Lorentzian and Euclidean $4$-simplexes were discussed, we briefly comment on the boundary data of a $4$-simplex in $\R^3$.
 
This case is a $4$-simplex with a metric of signature $0+++$ and has the same metric geometry as a linear immersion of the simplex into $\R^3$. Thus it follows that the map from the $4$-simplex boundary to $\R^3$ has to be orientation-reversing on some simplexes. A geometric way to see this is to imagine a $4$-simplex in 4d and gradually flatten it. Some of the outward normals to the tetrahedral faces of the $4$-simplex will point into the northern hemisphere, while others go to the southern hemisphere. As the boundary orientation plus the outward-facing normal agree with the standard orientation on $\R^4$, the tetrahedra with southward-pointing normals will be orientation-reversed with respect to ones with northward-pointing normals. Let us suppose that the standard orientation in $\R^3$ agrees with the tetrahedra with southward-pointing normals. Then applying the inversion map $\R^3\to\R^3$
$$I\colon \mathbf x\mapsto -\mathbf x,$$
to all the tetrahedra with northward-pointing normals gives all of the boundary tetrahedra embedded in 3d Euclidean space with orientation-preserving maps. This embedding of the tetrahedra in $\R^3$ defines a set of outward normal vectors to the triangular faces of the tetrahedra $\nb_{ab}$, and hence a set of boundary data for this geometry. Moreover the gluing maps $g_{ba}$ have a straightforward description with this representation. If two tetrahedra are both northern hemisphere or both southern hemisphere, then $g_{ba}$ is a spin lift of the identity rotation. If one is northern and one southern, then $g_{ba}$ is a spin lift of the rotation by $\pi$ about the normal to the common triangle in each tetrahedron, $\nb_{ab}=-\nb_{ba}$. A solution to the critical equations \eqref{equations} is given by $X_a=1\!\!1$.
The boundary data just described is not the only possible set for this geometry. All other sets of boundary data for this geometry are obtained by independently rotating each tetrahedron. This defines boundary data in the 3d Euclidean case. 
 
We are now ready to extend the classification and geometry theorems of \cite{Barrett2009}.

\begin{theo} \label{classification} Let the boundary data of a 4-simplex be geometric. Then
the critical point equations \eqref{equations} have 2 equivalence classes of solutions in the case of 4d Euclidean boundary data, 1 equivalence class in the case of 3d Euclidean boundary data, and no solutions in the case of 4d Lorentzian boundary data.
\end{theo}

{\em Proof.} Part of this theorem is proved in \cite{Barrett2009}. Specifically, it is shown there that there are 2 equivalence classes of solutions if any only if the boundary data is 4d Euclidean. 

The remainder of the theorem is proved in the rest of this section by giving a general procedure that operates on a given solution $\{X_a\}$ to give a second solution $\{X'_a\}$. The construction proceeds as follows.

A solution $\{X_a\}$ to the critical point equations can be written alternatively using the variables
$$X_{ab} = X_a^{-1} X_b,$$
which satisfy 
$$X_{ab}X_{bc}X_{ca}=1\!\!1 ,$$
for all $a,b,c$. 
The critical point equations \eqref{equations} then become
$$
X_{ab} \, \nb_{ba} = - \nb_{ab}.
$$
The involution which generates a second solution  $\{X'_{ab}\}_{a \neq b}$ of twenty group elements is constructed from the gluing maps and the first solution as follows
\be
\label{second}
X'_{ab} = g_{ab} \, X_{ba} \, g_{ab}.
\ee
In fact, since $X'_{ba} = (X'_{ab})^{-1}$ by construction, we have ten group elements $\{X'_{ab}\}_{a < b}$ satisfying the equations
$$
X'_{ab} \, \nb_{ba} = - \nb_{ab}.
$$
The set of group elements $\{X'_{ab}\}_{a < b}$ are however not solutions to the critical point equations unless each $X'_{ab}$ can be expressed as $X'_{ab} = (X'_a)^{-1} X'_b$ for some $X'_a$ and $X'_b$ in $\SU(2)$. This is equivalent to proving a cocycle condition for each face $abc$ of the complex dual to the $4$-simplex,
$$
X'_{ab} X'_{bc} X'_{ca} = 1 \!\! 1, 
$$
because the fundamental group of the three-sphere is trivial, and so all flat connections are pure gauge.
To show that the above cocycle condition is indeed satisfied we need the following lemma.

\begin{lem}\label{lem1}
Let $A,B,C$ be three $\SU(2)$ matrices corresponding to three $\SO(3)$ elements with rotation axes lying in the same plane $P$. Let $R$ be an $\SU(2)$ element associated to a rotation in the plane $P$. 
If $ABC = R$ then $CBA = R^{-1}$.
\end{lem}

{\em Proof.} Let $X$ be a $\pi$-rotation in the plane $P$. Conjugating both sides of the equality $ABC = R$ by $X$ leads to
$$
X A X^{-1} \, X B X^{-1} X C X^{-1} = X R X^{-1}.
$$
Now, it is immediate to see that $X D X^{-1} = D^{-1}$, for $D=A,B,C$, and $X R X^{-1} = R$. 
\vspace{-7.8mm}
\begin{flushright} $\square$ \end{flushright}

The above lemma is applied to our framework by defining the following three $\SU(2)$ matrices
\be
A = g_{ab} X_{ba}, \;\;\;\; B = g_{ab} g_{bc} X_{cb} \, g_{ba}, \;\;\;\; \mbox{and} \;\;\;\; C = g_{ab} g_{bc} g_{ca} X_{ac} \, g_{cb} g_{ba}.
\ee
These matrices satisfy $CBA = D_{abc}$, with $D_{abc} := g_{ab} g_{bc} g_{ca}$. Furthermore, using \eqref{equations} and \eqref{gluevector}, it is straightforward to check that the following three vectors
$$
\mathbf{s}_A = \nb_{ab}, \;\;\;\; {\mathbf s}_B  = g_{ab} \nb_{bc}, \;\;\;\; \mbox{and} \;\;\;\; {\mathbf s}_C = g_{ab} g_{bc} \nb_{ca},
$$
are stabilised by $A$, $B$ and $C$ respectively. This implies that the axes of the rotations associated to $A$, $B$ and $C$ are given by $\mathbf{s}_A$, $\mathbf{s}_B$ and $\mathbf{s}_C$ respectively. 

Now, recall that the $g_{ba}$ map corresponds to the parallel transport operator from tetrahedron $a$ to tetrahedron $b$ ; it encodes the change of reference frame from $a$ to $b$. 
This leads to two conclusions. 
Firstly, the $\SU(2)$ element $D_{abc} = g_{ab} g_{bc} g_{ca}$, mapping the tangent space of the tetrahedron $a$ to itself, corresponds to a rotation in the plane $P$ orthogonal to the edge $(abc)$ (hinge) shared by the three triangles $ab$, $bc$ and $ca$. This rotation stabilises the corresponding edge vector $\mathbf{e}_a = \mathbf x_a^v - \mathbf x_a^w$, constructed from the position vectors of the vertices $v,w$ bounding $(abc)$ in the frame associated to the simplex $a$, and the associated angle is the deficit angle around this edge. 
Secondly, because the normals $\nb_{ab}$, $\nb_{bc}$ and $\nb_{ca}$ correspond to three triangles sharing the $abc$ edge, they can all be defined in terms of the corresponding edge vector. Indeed, the normals can be constructed as follows
$$
\nb_{ab} = \mathbf{e}_a \times \mathbf{u}_{ab}, \;\;\;\; \nb_{bc} = \mathbf{e}_{b} \times \mathbf{u}_{bc}, \;\;\;\; \mbox{and} \;\;\;\; \nb_{ca} = \mathbf{e}_c \times \mathbf{u}_{ca}.
$$
where, the symbol `$\times$' denotes the 3d cross product, and $\mathbf{u}_{ab}$, $\mathbf{u}_{bc}$, $\mathbf{u}_{cd}$ are a second, distinct edge vector in each one of the triangles $ab$, $bc$ and $ca$ respectively, which is chosen such that $\nb_{ab}$ is pointing towards the exterior of simplex $a$. Now, $\mathbf{e}_a$, $\mathbf{e}_b$ and $\mathbf{e}_c$ are related by the parallel transport operator and we have that
$\mathbf{e}_b = g_{ba} \mathbf{e}_a$, and $\mathbf{e}_c = g_{cb} g_{ba} \mathbf{e}_a$. Therefore, 
one can readily see that 
$$
\nb_{ab} \cdot \mathbf{e}_a = 0, \;\;\;\; \nb_{bc} \cdot (g_{ba} \, \mathbf{e}_a) = 0, \;\;\;\; \mbox{and} \;\;\;\;  \nb_{ca} \cdot (g_{cb} \, g_{ba} \, \mathbf{e}_a) = 0,
$$
where the dot `$\cdot$' is the 3d Euclidean inner product. Using the invariance of the associated metric, that is,  $(g \mathbf x) \cdot \mathbf y = \mathbf x \cdot (g^{-1} \mathbf y)$, we can therefore see that
$\mathbf{s}_A$, $\mathbf{s}_B$ , and $\mathbf{s}_C$, the axes of rotation of the three matrices $A$, $B$ and $C$, all lie in the plane $\mathbf{e}_a^{\bot}$.

We can therefore apply lemma \ref{lem1} to the present context with the role of the matrix $R$ played by $D_{abc}$ and the plane $P = \mathbf{e}_a^{\bot}$ is given by the orthogonal complement of the edge $(abc)$. This leads to the conclusion that $ABC = D_{abc}^{-1}$ which reads, for each edge $(abc)$,
\be
X'_{ab} \, X'_{bc} \, X'_{ca} = 1\!\!1.
\ee
Therefore, for each triangle $(ab)$, there exits two $\SU(2)$ element $X_a'$ and $X_b'$ such that 
$$
X'_{ab} = (X'_{a})^{-1} X'_{b},
$$ 
and we have generated a second solution $\{X'_a\}$ to the critical point equations.
There are now two cases to consider. The two solutions $X$ and $X'$ are either inequivalent or equivalent.

\begin{itemize}

\item If the two solutions are inequivalent, the reconstruction theorem of \cite{Barrett2009} which states that the vector geometries determined by $\{X_a\}$ and $\{X'_a\}$ recombine into a single bivector geometry applies and the boundary data is 4d Euclidean.

\item It is also possible that the two solutions $X$ and $X'$ are in fact equivalent, that is, they are related by the symmetries \eqref{symmetry}, which implies that $X_{ab}'=\epsilon_a\epsilon_b X_{ab}$. Now if the boundary data for each tetrahedron is rotated so that the original solution is $X_a=1\!\!1$ for all $a$, then \eqref{second} implies that the gluing map satisfies $$g_{ab}^2=\pm1\!\!1$$ for each $a\ne b$. This means that the corresponding rotation squares to the identity in $\SO(3)$ and is thus either the identity element or a rotation by $\pi$ about the axis $n_{ab}$. To discuss the pattern of these two possibilities, define a parameter $\sigma_{ab}$ to take the value $-1$ for a rotation by $\pi$ and $+1$ for the identity rotation.

To show that this description of the boundary data and gluing maps $g_{ba}$ coincides with that of a 3d Euclidean geometry, it is necessary to show that the tetrahedra fall into two classes, with $\sigma_{ab}=-1$ for a transition between the two classes and $\sigma_{ab}=+1$ if both tetrahedra lie in the same class. Consider three distinct tetrahedra, labelled with $a$, $b$ and $c$. These tetrahedra have a common edge, which is represented by the displacement vector $\mathbf e_a$ in tetrahedron $a$ ($v$ and $w$ labelling its vertices, as above). Then
$$g_{ac}g_{cb}g_{ba}\mathbf e_a=\mathbf e_a. $$
But $g_{ba}\mathbf e_a=\sigma_{ab}\mathbf e_b$, etc., so that
$$\sigma_{ab}\sigma_{bc}\sigma_{ca}=1.$$
It follows that $$\sigma_{ab}=\sigma_a\sigma_b,$$
with $\sigma_a=\pm1$. Accordingly, the tetrahedra fall into two groups, those with $\sigma_a=+1$ and those with $\sigma_a=-1$, and $g_{ba}$ is a rotation by $\pi$ only when gluing one of one group with one of the other.
This shows that the geometry is exactly that of the 3d Euclidean case.

\end{itemize}
 
To complete the proof of theorem \ref{classification}, the argument is as follows. The explicit description of the 3d Euclidean boundary data shows that at least one equivalence class of solutions of \eqref{equations} exists for this case. There cannot be more than one because the part of the theorem proved in \cite{Barrett2009} would then show that boundary data is that of a non-degenerate 4d Euclidean 4-simplex. 

Conversely, if a solution to \eqref{equations} exists, then we obtain another one by applying the involution. By the explicit analysis of the case where the involution yields a symmetry-related solution we saw that this case corresponds to the boundary data of a 3d Euclidean 4-simplex. If it is inequivalent, the part of the theorem proved in \cite{Barrett2009} applies again, and it is 4d Euclidean. It follows immediately that for the Lorentzian case there are no solutions to these equations. 
\vspace{-7.1mm}
\begin{flushright} $\square$ \end{flushright}

\subsection{The action at the critical points}

In this section, we evaluate the action on solutions to the critical point equations and derive asymptotic formulae for the $\ftj$ symbol in the case of non-degenerate boundary data. 

\subsubsection{Non-geometric boundary data}

Lifting the critical point equation to coherent state space produces a phase
\be \label{lift}
(X_{a})^\dagger X_{b}| \mathbf{n}_{ba} \ra = e^{i \phi_{ab}} J|  \mathbf n_{ab} \ra,
\ee
for all $a\ne b$, from which follows that $\phi_{ab}=\phi_{ba}$.
This phase is the evaluation of the action as of course
$$
\bra{J \nb_{ab}}(X_{a})^\dagger X_{b}| \mathbf{n}_{ba} \ra^{2k_{ab}} = e^{2i  k_{ab} \phi_{ab}},
$$
and thus the action \eqref{action} evaluates to
\be\label{action-crit}
S_{(k,\nb)}|_{crit} = 2 i \sum_{a < b} k_{ab} \phi_{ab}.
\ee
When scaling up the $k$ this will provide exactly the phase of the oscillating part of the asymptotic formula. Note that the phase $\phi_{ab}$ depends on the arbitrary phase of the coherent states. While this arbitrary phase cancels in the full state sum we need to make a choice when decomposing the state sum into individual weights. 

Under the assumption that the boundary data is non-degenerate (although non-geometric), we have seen that there is at most one solution to the critical point equations \eqref{equations} up to equivalence. Thus, we can change the $\ket{\nb_{ab}}$ to $e^{i \phi_{ab}} \ket{\nb_{ab}}$ for $a<b$, and set the action at the critical point to zero. This choice of course depends on the full set $\{k_{ab}$, $\nb_{ab}\}$. Using this phase choice we can now easily evaluate the asymptotic behaviour of the $\ftj$ symbol
\be \label{su2asym-general}
\ftj(\lambda k,\nb) \sim (-1)^\chi \left( \frac{2 \pi}{ \lambda} \right)^6 \frac{2^4}{(4 \pi)^8} \frac{1}{\sqrt{\det H}},
\ee
where $H$ is the Hessian of the action \eqref{action} evaluated on the critical points and the square root is defined as in \cite{Barrett2009}.

\subsubsection{Geometric boundary data}

If the boundary data defines a positive definite metric on the boundary of the 4-simplex we have a geometrically distinguished boundary state, the Regge state, defined by \eqref{ReggeState}.
We can then look at the three cases of the classification theorem in turn with the boundary state given by a Regge state.

\paragraph{Two inequivalent solutions.}

Given two inequivalent solutions $b^\pm_{ab} = k_{ab} X^\pm_a \nb_{ab}$, we know from the reconstruction theorem in \cite{Barrett2009} that we have a pair of inversion-related non degenerate 4-simplices $\sigma$ with bivectors given by their self-dual and anti-self-dual parts by $B_{ab}(\sigma) = (b^+_{ab}, b^-_{ab})$, together with a pair with parity-related geometry with bivectors given by $B_{ab}(P\sigma) = - (b^-_{ab}, b^+_{ab})$. This notation fixes our choice of $\pm$ labels corresponding to the self-dual and anti-self-dual components of the bivectors for the two solutions.

From the lift \eqref{lift} we now obtain two set of angles $\phi_{ab}^\pm$. By substituting \eqref{second} and \eqref{ReggeState} into \eqref{lift} we have immediately that $$e^{i\phi^+_{ab}} = e^{- i\phi^-_{ab}}.$$ The more detailed analysis in \cite{Barrett2009} shows that these angles are, up to factors of $\pi$ that cancel once exponentiated, equal to plus or minus half the dihedral angle $\Theta_{ab}$ of the triangle shared by tetrahedra $a$ and $b$ $$\phi^\pm_{ab} = \pm \frac{1}{2}\Theta_{ab}.$$ As the $k_{ab}$ give the areas of the geometry the action thus evaluates to the Regge action $S_R (\sigma)$ for the $4$-simplex $\sigma$ determined by the boundary data
\be\label{action-geom}
S_{(k,\nb)}|_{\pm} = \pm i \sum_{a < b} k_{ab} \Theta_{ab} = \pm i S_R (\sigma).
\ee

The full asymptotic behaviour is then given by:
\be \label{su2asym-geom}
\ftj(\lambda k,\nb) \sim (-1)^\chi \left( \frac{2 \pi}{ \lambda} \right)^6 \frac{2^4}{(4 \pi)^8} \left( \frac{e^{i \lambda S_R (\sigma)}}{\sqrt{\det H_+}} + \frac{e^{- i \lambda S_R (\sigma)}}{\sqrt{\det H_-}}\right),
\ee
with $H_\pm$ the Hessian of the action evaluated at $b_\pm$ respectively.

Note that as we have two terms now an arbitrary phase change in the coherent states could at most set one of these phases to zero, $\phi^+ - \phi^- = \Theta$ is independent of the arbitrary phase. The geometric phase choice is significantly stronger than that in the non-geometric sector as it can be applied directly to the boundary data, without having to solve the critical point equations.

\paragraph{Single solution.}

In this case we have seen that we have the geometry of a degenerate 4-simplex in $\R^3$ and a solution is given by the identity rotations. Then the phases $\phi_{ab}$ evaluate to the phase of $g_{ba}$. The action is a degenerate case of \eqref{action-geom}, with $\Theta_{ab}=0$ or $\pi$. As per the discussion of the flat geometry we have $\Theta_{ab} = \pi$ exactly if $a$ and $b$ are oppositely oriented. Therefore in the action $\pi$ is multiplied by a sum of spins belonging to a closed surface and these always sum to an integer. Since the action is a multiple of $\pi$, the $\pm$ sign is irrelevant, and $e^{iS_R(\sigma)}=\pm1$.  The full asymptotic behaviour is given by this sign times \eqref{su2asym-general}.

\paragraph{No solutions.}

This situation occurs when the boundary geometry is that of a 4-simplex with Lorentzian metric. In this case, the $\ftj$ symbol goes to zero asymptotically faster than any polynomial of $\lambda$.

\section{Quantum gravity asymptotics from the $\ftj$ symbol}

In this section, we show how the previous results relate to models of Euclidean quantum gravity.

\subsection{Quantum gravity models as constrained squares of the Ooguri model}

All the new quantum gravity models are closely related to the Ooguri state sum which is constructed using the $\SU(2)$ $\ftj$ symbols. In fact most of them can be written as constrained versions of the square of the Ooguri model in the coherent state basis. The  state sum model is given by identifying the boundary data on each $4$-simplex appropriately and integrating over it. The boundary data for the gravity model on a 4-simplex is a set of unit vectors $\mathbf n_{ab}$ and a set of spins $j_{ab}$, in other words exactly the same data as for an $SU(2)$ $\ftj$ symbol.
The precise details of the gluing of simplexes are beyond the scope of this paper, but some further remarks about the full state sum are made in section \ref{discussion}.

There are two types of model given by products of $\ftj$ symbols. The first has 4-simplex amplitude given by
 rescaled squares of the $\ftj$ symbol
\be
Z(j,\nb) = \ftj(j_{ab},\nb_{ab}) \times \ftj(c_\gamma j_{ab},\nb_{ab}).
\ee
using the parameter $c_\gamma = \frac{|1-\gamma|}{1+\gamma}$.

This gives both the EPRL and FK models for parameter $\gamma<1$, and on setting $\gamma=0$ (i.e., $c_\gamma=1$), the EPR model.

The second type of model has 4-simplex amplitude given by
  \be
Z^{cc}(j,\nb) = \ftj(j_{ab},\nb_{ab}) \times \overline{\ftj(c_\gamma j_{ab},\nb_{ab})}.
\ee
and includes the FK model with parameter $\gamma>1$, and for $\gamma=0$, the FK model without a parameter. 
The only model that can not be written in this form is the EPRL model with $\gamma > 1$ which takes the form $Z^{cc}$  asymptotically. 

Note that these models are usually written as function of $k$ such that $j =\frac{1+\gamma}{2} k$ and $j' =\frac{|1-\gamma|}{2} k$. The possible values of $\gamma$ and $j$, $j'$ and thus $k$ are restricted as the state sum is identically zero unless $c_\gamma j$ takes on half integer values.

This formulation was the way the FK model \cite{Freidel2008} was initially written down, and for the EPRL model this formulation was developed in \cite{Livine2007a,Barrett2009}.

\subsection{Quantum gravity $4$-simplex asymptotics from the $\SU(2)$ $\ftj$ symbol}

The $4$-simplex amplitudes for the models without complex conjugation are then given by
\be
Z(j,\nb) = (-1)^{\chi'} \int_{\SU(2)^{10}} \prod_{a=1...5} dX_a dX'_a \, \bra{J \nb_{ab}} X_a^\dagger X_b \ket{\nb_{ba}}^{2 j_{ab}} \bra{J \nb_{ab}} {X'}_a^\dagger X'_b \ket{\nb_{ba}}^{2 c_\gamma j_{ab}}.
\ee
For the models with complex conjugation, the $4$-simplex amplitudes read
\be
Z^{cc}(j,\nb) = (-1)^{\chi'} \int_{\SU(2)^{10}} \prod_{a=1..5} dX_a dX'_a \, \bra{J \nb_{ab}} X_a^\dagger X_b \ket{\nb_{ba}}^{2 j_{ab}} \overline{\bra{J \nb_{ab}} {X'}_a^\dagger X'_b \ket{\nb_{ba}}}^{\, \,2 c_\gamma j_{ab}}.
\ee
The corresponding actions can be written easily in terms of the $\ftj$ action \eqref{action}:
\bea
S^\gamma_{(j,\nb)}[X, X'] &=& S_{(j,\nb)}[X] + c_\gamma S_{(j,\nb)}[X'],\nn\\
S^{\gamma,cc}_{(j,\nb)}[X, X'] &=& S_{(j,\nb)}[X] + c_\gamma \overline{S_{(j,\nb)}[X']}.
\eea
The equations governing the stationary phase analysis of all these models are identical and are simply \eqref{closure} and two copies of \eqref{equations}:
\bea \label{equations2}
&\displaystyle\sum_{b: b\ne a} j_{ab}\nb_{ab}=0,&\nn\\
&X_b \nb_{ba} = - X_a\nb_{ab},&\nn\\
&X'_b \nb_{ba} = - X'_a\nb_{ab}.&
\eea
We can thus directly apply the geometric discussion above to the solutions to these equations in terms of $b_{ab} = j_{ab} X_{a} n_{ab}$. In particular it is clear that given any set $\{b_{ab}, X_a \}$ solving the $\SU(2)$ equations we obtain a solution of the equations for the new models simply by setting $X'_a = X_a$. Generally, the asymptotic behaviour can be obtained by simply multiplying the asymptotics of the two $\SU(2)$ factors.

\subsubsection{Non-geometric boundary data}

We obtain two angles, $\phi_{ab}$ and $\phi'_{ab}$ from the lift of \eqref{equations2}. These combine to give the summands in the action of the models without complex conjugation
\be \label{phases}
\bra{J \nb_{ab}}(X_{a})^\dagger X_{b}| \mathbf{n}_{ba} \ra^{2 j_{ab}}\bra{J \nb_{ab}}(X'_{a})^\dagger X'_{b}| \mathbf{n}_{ba} \ra^{2 c_\gamma  j_{ab}} = e^{i 2 j_{ab} (\phi_{ab} + c_\gamma \phi'_{ab})},
\ee
Analogously for the models with conjugation the phase is $ \phi_{ab} - c_\gamma\phi'_{ab}$. Thus the actions evaluate to
\bea
S^\gamma_{(j,\nb)}|_{crit} &=& 2 i \sum_{a < b} j_{ab} (\phi_{ab} + c_\gamma \phi'_{ab}),\nn\\
S^{\gamma,cc}_{(j,\nb)}|_{crit} &=& 2 i \sum_{a < b} j_{ab} (\phi_{ab} - c_\gamma \phi'_{ab}).
\eea 
 
Restricting to non-geometric boundary data which are non-degenerate, we know that there is one or no solution to the critical point equations \eqref{equations}, up to equivalence. Thus we have that $\{X_a\} \sim \{X'_a\}$ and therefore that $\phi_{ab} = \phi'_{ab}$. Thus, just as in the case of a single $\SU(2)$ we can now set the phase in $\ket{\nb_{ab}}$ such that it cancels the $\phi_{ab}$ to obtain the final asymptotic formula by multiplying two copies of \eqref{su2asym-general} scaled by $c_\gamma$
\be \label{modelasym-gen}
Z(\lambda j,\nb) \sim (-1)^\chi \left( \frac{2 \pi}{ \lambda} \right)^{12} \frac{2^8}{(4 \pi)^{16}} \frac{1}{{c_\gamma}^6|\det H|},
\ee
where $H$ is the same Hessian of the action \eqref{action} evaluated on the critical points with spin given by $j$. As the Hessian is linear in the $j$ and 12 dimensional, the determinant evaluated at the critical point $j' = c_\gamma j$ differs by ${c_\gamma}^{12}$ from that evaluated at $j$. This accounts for the factor ${c_\gamma}^6$.

\subsubsection{Geometric boundary data}

If the boundary data defines a positive definite $3$-metric on the boundary of the 4-simplex, we again chose the boundary state to be a Regge state \eqref{ReggeState}, and the classification proceeds in the same way as before.

\paragraph{Two inequivalent solutions.}

As before given two inequivalent solutions $b^\pm_{ab} = j_{ab} X_a^\pm \nb_{ab}$ they necessarily correspond to the self-dual and anti self-dual parts of geometric 4-simplex bivectors given by $B_{ab}(\sigma) = (b^+_{ab}, b^-_{ab})$ and $B_{ab}(P\sigma) = - (b^-_{ab}, b^+_{ab})$.

These combine to four solutions to the stationary and critical point equations \eqref{equations2}:
$$(b, {c_\gamma}^{-1} \, b') \in \{(b^+,b^+),(b^+,b^-),(b^-,b^+),(b^-,b^-)\}.$$
The $++$ and $--$ solutions are analogous to the solutions for non-geometric boundary data as the solution is just the double of a pure $\SU(2)$ solution. With the Regge boundary state the phase in \eqref{lift} and thus the phases $\phi$, $\phi'$ in \eqref{phases} evaluate to $\phi_{ab}|_{b^\pm} = \pm \frac{1}{2} \Theta_{ab}$.

The full asymptotics is then again given in terms of the Regge action $S_R(\sigma)$ for the $4$-simplex $\sigma$ determined by the boundary data by a sum over the signs of the 4 critical points $(b^\pm,b^\pm)$. That is, with $\epsilon, \epsilon' = \pm 1$,
\be
Z(\lambda j,\nb) \sim (-1)^\chi \left( \frac{2 \pi}{ \lambda} \right)^{12} \frac{2^8}{(4 \pi)^{16}} \sum_{\epsilon, \epsilon' = \pm 1} \frac{e^{i \lambda (\epsilon + \epsilon' c_\gamma) S_R(\sigma)}}{{c_\gamma}^6\sqrt{\det H_\epsilon \det H_{\epsilon'}}},
\ee
for the models without complex conjugation and by
\be
Z^{cc}(\lambda j,\nb) \sim (-1)^\chi \left( \frac{2 \pi}{ \lambda} \right)^{12} \frac{2^8}{(4 \pi)^{16}} \sum_{\epsilon, \epsilon'=\pm 1} \frac{e^{i \lambda (\epsilon - \epsilon' c_\gamma)  S_R(\sigma)}}{{c_\gamma}^6\sqrt{\det H_\epsilon \overline{\det H_{\epsilon'}}}},
\ee
for those with. As the amplitudes leading to these are rescaled squares of the $\SU(2)$ $\ftj$ symbol this is of course just the rescaled square of the $\SU(2)$ asymptotics \eqref{su2asym-geom}.

We can rewrite the actions further by expressing them in terms of $k_{ab} = \frac{2}{(1+\gamma)} j_{ab}$. Writing $S_{\epsilon \epsilon'}$ for $S|_{(b^\epsilon, b^\epsilon)}$ we obtain 
\bea
S^\gamma_{\epsilon \epsilon'} &=& (\epsilon(1+\gamma) + \epsilon' |1-\gamma|) \frac{1}{2} \sum_{a<b} k_{ab} \Theta_{ab},\nn\\
S^{\gamma,cc}_{\epsilon \epsilon'} &=& (\epsilon (1+\gamma) - \epsilon' |1-\gamma|) \frac{1}{2} \sum_{a<b} k_{ab} \Theta_{ab},\nn
\eea
and thus 
\be
S^{\gamma > 1}_{\epsilon \epsilon'} = S^{\gamma < 1, cc}_{\epsilon \epsilon'} = ((\epsilon + \epsilon')\gamma + (\epsilon - \epsilon')) \frac{1}{2} \sum_{a<b} k_{ab} \Theta_{ab},\nn
\ee
\be\label{nicemodels}
S^{\gamma > 1, cc}_{\epsilon \epsilon'} = S^{\gamma < 1}_{\epsilon \epsilon'} = ((\epsilon - \epsilon')\gamma + (\epsilon + \epsilon')) \frac{1}{2} \sum_{a<b} k_{ab} \Theta_{ab},
\ee
Hence, we have in particular
$$S_{\mbox{{\tiny EPR}}}|_{\epsilon \epsilon'} = S^{\gamma<1}|_{\epsilon \epsilon' \, \gamma = 0} = \frac{\epsilon + \epsilon'}{2} \sum_{a<b} k_{ab} \Theta_{ab},$$
and 
$$S_{\mbox{{\tiny FK}}}|_{\epsilon \epsilon'} = S^{cc, \gamma<1}|_{\epsilon \epsilon'\, \gamma = 0} = \frac{\epsilon - \epsilon'}{2} \sum_{a<b} k_{ab} \Theta_{ab}.$$
This is the form in which the actions appeared in the literature so far, where the state sums were expressed in terms of $k$.

\paragraph{Single solution.}
As in the $\SU(2)$ BF case the action is zero or $\pi$ and the amplitude evaluates to \eqref{modelasym-gen} times a sign.

\paragraph{No solutions.}
In this case the boundary geometry is that of a 4-simplex with Lorentzian metric and we again have that the amplitude is suppressed exponentially for large spins.

\section{Discussion}\label{discussion}

The $\SU(2)$ $\ftj$-symbol is the building block for the Ooguri model which is a topological model based on flat $\SU(2)$ connections. Although it is much simpler than a quantum theory of gravity, the results of this paper have demonstrated that understanding the $\ftj$-symbol and its asymptotics is the key to understanding the more complicated gravity models. Curiously, the 
$\ftj$-symbol exhibits some gravity-like features itself.

There is a heuristic dictionary which translates between the discrete variables and discrete action entering into asymptotics formulae, and the traditional continuum fields and their actions constructed from Lagrangians. This dictionary has been started in the main text, where the interpretation of the general asymptotic data for an  $\SU(2)$ $\ftj$-symbol is interpreted as  discrete $B$ and $A$ fields. This analogy goes further as it is possible to construct the Ooguri model from a $BF$ Lagrangian, in much the same way as the Ponzano-Regge model is related to a $BF$ theory in three dimensions.
In fact, a configuration of the asymptotic data for an  $\SU(2)$ $\ftj$-symbol is a vector geometry which gives a constant $\su(2)$-valued $B$ field on the 4-simplex and a discrete $\SU(2)$ connection determined by the $X$ matrices. In the context of a state sum model on a manifold it is the sum over the spin labels $k$ on each triangle that enforces the flatness of the holonomy around the triangle \cite{Ooguri1992c}. This flatness is not visible by examining a single 4-simplex.

The analogy goes further. A $B$ field is `geometric' if it is the self-dual part of $e\wedge e$ for some non-degenerate frame field $e^I$
\begin{equation}B^i=P^i_{IJ}e^I\wedge e^J,\label{geometricB}\end{equation}
with $P^i_{IJ}$ projecting antisymmetric $4\times4$ matrices to their self-dual part, indexed by $i$.
If $B$ here is the constant $\su(2)$-valued 2-form on a 4-simplex determined by a critical point, it satisfies this equation precisely when the boundary data is geometric. This is because a geometric boundary induces a constant metric on the 4-simplex, for which a frame field exists. 
One can see that there are five constraints involved in these equations. The constant $B$ field has 18 components, whereas the geometric ones form a space of dimension 13. (The $e$ field has 16 components but three-parameter's worth are identified in \eqref{geometricB}, due to the invariance of the self-dual part under anti-self-dual rotations.)

A further confirmation of this interpretation is given by the continuum analogue of \eqref{second}.
In this equation, $X$ represents the self-dual part of the connection $\omega_+$, $X'$ the anti-self-dual part $\omega_-$ and $g$ the Levi-Civita connection on the boundary $\omega_0$. Linearising  \eqref{second}
leads to
$$\omega_-= 2\omega_0-\omega_+,$$
which is the right relation for these connection forms restricted to the boundary if the two-form $B$ is geometric.

This dictionary also gives a heuristic explanation for the appearance of the Regge action for geometric boundary data in the $\ftj$-symbol. This is because in the continuum picture, the $\SU(2)$ $BF$ action is exactly the action for self-dual gravity on substituting with \eqref{geometricB}, as discovered by Plebanski. 

Hence the restriction to geometric or Regge-like boundary data is the precise counterpart in the discrete setting of equation \eqref{geometricB} in a continuum Lagrangian. This suggests that it is possible to define a discrete version of Plebanski's gravity, in which constraints are added to a $BF$ Lagrangian to enforce \eqref{geometricB}. The quantisation with a state sum model is given by the usual formula (here somewhat simplified)
$$Z_\text{PL}=\int_{\{\mathbf n\}}\sum_k\prod_\text{4-simplexes} \ftj\text{-symbol}$$
but with the integral over the coherent states restricted to ones that specify geometric boundary data on each 4-simplex. Whether such a model is simple enough or useful remains to be seen. More generally, one could seek modifications of the Ooguri model so that the geometric configurations dominate and the Plebanski Lagrangian, and hence general relativity, is evident in an asymptotic regime. Such a modification would of course have to break the reduction of the Ooguri model to flat connections.


The apparent paradox that the topologically trivial Ooguri model contains gravity in its asymptotics is resolved by the fact the sum over the spins in the Ooguri state sum model will always enforce flatness. Thus it is possible to have geometric boundary data for an arbitrary triangulation of a 4-ball but the asymptotics will always be determined by the Einstein action for flat metrics, as in the three-dimensional case \cite{Dowdall2009}.

Moving now to the discussion of the gravity state sum models based on $\SU(2)\times\SU(2)$, the results here demonstrate that the asymptotics of these models can be explained by a suitable squaring of the  $\SU(2)$ $\ftj$ symbol asymptotics. This gives a very nice explanation of the four terms in the asymptotics found
for the EPRL/FK models. These are just obtained from squaring an asymptotic formula with two terms. In particular this demonstrates that the two of these terms that were unexpected (called the `weird terms' in \cite{Barrett2009}) are definitely required.

However the analysis of this result, which was the original motivation for the paper, delivered also another insight into the gravity models: the asymptotic formula \emph{always} contains the terms which correspond to those of a \emph{single} $\SU(2)$  Ooguri model for the same boundary data $(k,\mathbf n)$. This is because critical points of the gravity model of the form $X^+=X^-$ just correspond to critical points of the $\SU(2)$  $\ftj$ symbol, and, moreover, the action of the gravity models \eqref{nicemodels} reduces to the $\SU(2)$  action \eqref{action-geom} on setting $\epsilon=\epsilon'$. The Hessians may differ, however. In a sense this explains all of the asymptotic terms except the genuine gravity terms (the genuine gravity terms being those with asymptotics given by $\gamma$ times the Regge action, for which $X^+$ and $X^-$ are inequivalent). A similar conclusion holds for the Lorentzian model based on the representations of $\SL(2,\C)$. Apart from the expected gravity terms given by $\gamma$ times the Lorentzian Regge action, all of the other asymptotic terms correspond to the $\SU(2)$ Ooguri model.
 
This picture can be understood heuristically from the Lagrangian point of view. The gravity models are obtained by constraining a $BF$ theory and so the Lagrangian should be a generalisation of the Reisenberger Lagrangian \cite{Reisenberger:1998fk,Buffenoir:2004vx}, which corresponds to our case with $c_\gamma=1$. The Reisenberger Lagrangian is the $\SO(4)$ or $\SO(3,1)$  $BF$ Lagrangian with a constraint term added which enforces the equality of the area metric from the self-dual part with the area metric from the anti-self-dual part of the $B$ field. 
One class of solutions is given by
 $$B^{IJ}=\frac12 \epsilon^{IJ}_{KL}e^K\wedge e^L,$$
  which correspond to our `geometric' boundary data and the gravity asymptotic terms. A second class of solutions is given by an $\SU(2)$ connection $A$ and the matrix $ B^{IJ}$ lying in the $\su(2)$ Lie algebra, for which the action is the $\SU(2)$ $BF$ action with no constraints. This class of solutions corresponds to our non-geometric ones. 

Finally, some remarks about the full state sum models and their critical points. The question of whether the Regge equations of motion appear correctly from the state sum models is an important question. From the results developed here, it is possible to draw some preliminary conclusions. In the case of the Euclidean models, geometric boundary data for a 4-simplex is always embedded in a larger space of non-geometric boundary data. This can be seen from the fact that the vectors $\mathbf b_{ab}$ span an 18-dimensional vector space but the geometric ones form a submanifold of dimension only 13. Thus one can always perturb the $\mathbf b_{ab}$ away from the geometric confiurations, and hence one can perturb the boundary data given by the $\mathbf n_{ab}$ away from the geometric configuations too. Thus at first sight it seems that the variations are more than the variations of the metric geometry in Regge calculus, and hence one will get more equations of motion. However the fact that the number of solutions drops immediately on leaving the space of geometric solutions means that the situation is a little more complicated; there is interference between the geometric sector and the $\SU(2)$  $BF$ sector. If one could separate out the geometric solutions, then these would satisfy the Regge equations, but the $\SU(2)$  $BF$ ones certainly do not. More work is required to determine the status of the Regge equations in this model.

For the Lorentzian theory however, the situation is much simpler and clearer. The gravity solutions are well separated from the  
$\SU(2)$  $BF$ solutions and all though the state sum model contains both in its asymptotics, perturbing one does not give the other. Thus the Lorentzian gravity configurations can only be perturbed to other Lorentzian gravity configurations. This means that the correct Regge-like gluing of 4-simplexes is preserved under perturbations. Thus perturbations can lead only to the Regge equations of motion. Thus the state sum model does indeed contain asymptotic terms which approximate general relativity.

\section{Acknowledgments}

We thank Bianca Dittrich for discussions on the gluing constraints. WF acknowledges funding from the Royal Commission for the Exhibition of 1851.

\newcommand{\etalchar}[1]{$^{#1}$}

\end{document}